\documentclass[12pt]{article}
\usepackage[utf8]{inputenc}

\usepackage{geometry}

\usepackage{natbib}
\usepackage{amsmath}
\usepackage{amssymb}
\usepackage{xcolor}
\usepackage{hyperref}
\usepackage{booktabs}
\usepackage{amsthm}
\usepackage{bm}
\usepackage{pdflscape}
\usepackage{setspace}
\usepackage{graphicx}
\usepackage{subcaption}
\usepackage{soul}
\usepackage{placeins}
\usepackage{float}

\allowdisplaybreaks

\hypersetup{colorlinks=true,linkcolor=blue,filecolor=blue,citecolor=blue}

\usepackage{tikz}
\usetikzlibrary{arrows.meta}
\usepackage[labelfont=bf]{caption}
\usepackage{relsize}
\usepackage{xifthen}
\usepackage{amsmath,amssymb,xcolor}
\usepackage{pgf,tikz}
\usetikzlibrary{arrows,shapes.arrows,shapes.geometric,
shapes.multipart,decorations.pathmorphing,positioning,
swigs}

\numberwithin{intassumption}{assumption}

\newcommand{\E}{\mathbb{E}}
\newcommand{\PP}{\mathbb{P}}
\newcommand{\bX}{\boldsymbol X}

\newcommand{\indep}{\perp \!\!\! \perp}

\newcommand{\sumn}{\frac{1}{n}\sum_{i=1}^n}

\newcommand{\missingnessdag}[1]{%
  \begin{tikzpicture}[
    >=Stealth,
    scale=0.7,
    transform shape,
    every node/.style={font=\large},
    every path/.style={thick}
  ]
    \path[use as bounding box] (-0.3,-1.6) rectangle (3.5,1.0);

    \node (X) at (0,0) {$X$};
    \node (A) at (1.6,0) {$A$};
    \node (Y) at (3.2,0) {$Y$};
    \node (R) at (3.2,-1.3) {$R$};

    \draw[->] (X) -- (A);
    \draw[->] (X) to[bend left=35] (Y);

    #1
  \end{tikzpicture}%
}

\title{Addressing Missing Exposures in Causal Inference: A Practical Guide}
\author{Mo Wang, Elizabeth A. Stuart, Keith Barnatchez
\\
{\small Department of Biostatistics, Johns Hopkins Bloomberg School of Public Health}}

\begin{document}

\maketitle

\begin{abstract}
Missing exposure data are a common challenge in epidemiologic studies that can complicate the identification and estimation of causal effects. Particularly for researchers constructing doubly-robust estimators, it can be unclear how to adapt standard methods such as augmented inverse probability weighting (AIPW) to account for missing exposures. This paper clarifies  challenges posed by missing exposures, and discusses approaches available for doubly-robust estimation in these settings. We outline numerous mechanisms that can give rise to missing exposures in practice, and demonstrate why double-robustness alone  does not protect complete-case AIPW from missing exposure bias. We highlight three broad approaches for addressing this bias: multiply-robust estimation,  single imputation and multiple imputation. Using complete-data AIPW---the estimator one would construct in the absence of missing data---as a benchmark, we demonstrate how these approaches impute different components of the ideal complete-data estimator. We discuss practical implementation challenges, including whether outcome information should be used to impute exposures or predict exposure missingness. Through simulations, we illustrate the critical roles of the missingness mechanism, treatment effect heterogeneity and model specification on the performance of these approaches. We conclude with recommendations to help researchers choose an analysis strategy and inform modeling choices involved in their implementation.
\end{abstract}
\textit{Key words}: missing exposures, causal inference, augmented inverse probability weighting, multiple imputation

\doublespacing

\thispagestyle{empty}
\setcounter{page}{0}
\clearpage
\section{Introduction}
\label{sec:introduction}

Missing exposure data are common  in epidemiological and public health research, and can emerge in a broad range of applications.  For example, in studies using electronic health records (EHRs), exposure status defined by a baseline laboratory measurement may be unavailable because the test was not ordered or its result was not recorded \citep{TanEtAl2023}. In cohort studies, information on behavioral exposures, such as physical activity, may be missing when participants miss a study visit or do not complete the relevant exposure assessment \citep{ShortreedForbes2010}. In environmental health studies, exposure measurements may be missing because monitoring systems record pollutant concentrations intermittently rather than continuously \citep{SamoliEtAl2014}.

For researchers aiming to estimate causal effects in the above settings, missing exposures can considerably complicate the validity and implementation of otherwise familiar methods. In the absence of missing data, doubly-robust methods such as augmented inverse probability weighting (AIPW) have emerged as  increasingly popular estimation procedures, owed in part to their robustness to model misspecification and ability to accommodate machine learning when modeling components of the underlying data. For researchers familiar with AIPW, it can be unclear how to adapt the standard estimation procedure when exposures are missing. Dropping units with incomplete exposures can induce selection bias and reduce efficiency \citep{ZhangEtAl2016}, while imputing missing exposures introduces additional modeling requirements and a need to appropriately propagate uncertainty from imputation \citep{DondersEtAl2006}. 

Several approaches are available for handling \textit{general} forms of missing data, including complete-case analysis, single and multiple imputation, and multiply-robust methods that combine inverse probability weighting and regression models. These approaches provide potential ways to adapt AIPW in the presence of missing exposures, but their validity and relative merits strongly depend on the underlying observation mechanism. Critically, missing exposures  introduce unique complications relative to settings where only outcomes are missing: it can be unclear whether models used to address exposure missingness should include only baseline covariates, or also the observed outcome. This confusion partly arises from the general emphasis in causal inference on separating design and analysis in propensity score methods \citep{rubin2007design}, which would seem to imply that outcome information should be excluded when imputing exposures or predicting the probability of missingness. It can be unclear whether this principle extends to models used to account for missing exposures, particularly if the outcome itself is predictive of exposure availability \citep{MoonsEtAl2006}. These challenges leave researchers with uncertainty in both how to implement each approach and compare their relative merits.

This paper examines the consequences of missing exposure data under a variety of situations and introduces common approaches to estimating the average treatment effect when the exposure is partially missing.  Building on prior work on doubly robust estimation with missing exposures \citep{williamson2012doubly}, we clarify connections between imputation-based and multiply-robust approaches by showing how each approach effectively reconstructs the ideal AIPW estimator one would construct with complete data.  In our simulation study, we compare complete-case augmented inverse probability weighting (AIPW), single and multiple imputation followed by AIPW, and a multiply-robust (MR) estimator for missing exposure data. The examples illustrate how exposure observation mechanisms and nuisance model specifications affect bias, root mean squared error (RMSE), and confidence interval coverage, with the aim of supporting informed choices in epidemiological applications.


\section{Complications Arising  with Missing Exposures}
\label{sec:missing-exp}

\subsection{Notation}

We begin by introducing notation under two scenarios. In the first, ideal scenario, we observe $n$ independent samples of  the \textit{complete}  unit $O^\text{C}=(\bX,A,Y)$, where $A$ is a binary exposure, $\bX$ collects baseline covariates, and $Y$ is the outcome. In the second scenario, we instead observe $n$ independent samples of $O=(\bX,R,RA,Y)$. $R$ is a binary indicator of whether the exposure is observed,  so that $A$ is available for units with $R=1$ and missing otherwise. Throughout this paper, we use the term ``exposure"  to broadly refer to a variable whose causal effect on an outcome is of interest. This may be a medical treatment, a behavior, or an environmental exposure.

Let $Y(a)$ denote the potential outcome experienced under exposure level $a\in\{0,1\}$.  Our causal estimand is the average causal effect of exposure , 
conventionally termed the average treatment effect (ATE):
\[
    \tau=\mathbb{E}\{Y(1)-Y(0)\}.
\] 
In the ideal scenario where we have access to complete data $O^\text{C}$, one typically proceeds under a set of ``core" causal assumptions commonly invoked in epidemiologic studies: \textbf{(1)} consistency $Y = AY(1) + (1-A)Y(0)$; \textbf{(2)} positivity of exposure status $0 < \PP(A=1\mid \bX) < 1$; and \textbf{(3)} conditional exchangeability of exposure assignment $(Y(1), Y(0)) \indep A \mid \bX$.  Under these assumptions, in the absence of missing exposures where we observe $O^\text{C}$ it is  well-established that  $\tau$ is identified by both the G-formula and inverse probability weighting (IPW) estimands \citep{Hernan2025CausalInference}
\[
\tau = \E[m(\bX,1) - m(\bX,0)] = \E\left[ \left(\frac{A}{g(\bX)} - \frac{1-A}{1-g(\bX)}\right)Y\right],
\]
where $m(\bX,a) = \E(Y\mid A=a,\bX)$ is the outcome regression function conditional on exposure and baseline covariates, and $g(\bX) = \PP(A=1 \mid \bX)$ is the exposure propensity score function. Unfortunately, when exposures are partially missing so that we only observe $O$, the G-formula and IPW estimands themselves are not identified without further assumptions.

\begin{figure}[t!]
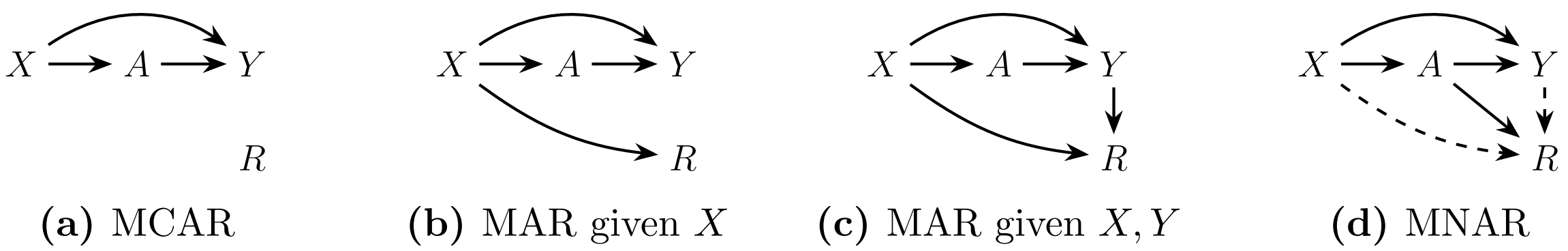

  \centering
  \begin{subfigure}[t]{0.24\textwidth}
    \centering
    \missingnessdag{
      \draw[->] (A) -- (Y);
    }
    \caption{MCAR}
    \label{fig:dag-mcar}
  \end{subfigure}\hfill
  \begin{subfigure}[t]{0.24\textwidth}
    \centering
    \missingnessdag{
      \draw[->] (A) -- (Y);
      \draw[->] (X) to[bend right=15] (R);
    }
    \caption{MAR given $X$}
    \label{fig:dag-mar-x}
  \end{subfigure}\hfill
  \begin{subfigure}[t]{0.24\textwidth}
    \centering
    \missingnessdag{
      \draw[->] (A) -- (Y);
      \draw[->] (X) to[bend right=15] (R);
      \draw[->] (Y) -- (R);
    }
    \caption{MAR given $X,Y$}
    \label{fig:dag-mar-xy}
  \end{subfigure}\hfill
  \begin{subfigure}[t]{0.24\textwidth}
    \centering
    \missingnessdag{
      \draw[->] (A) -- (Y);
      \draw[->] (A) -- (R);
      \draw[->, dashed] (X) to[bend right=15] (R);
      \draw[->, dashed] (Y) -- (R);
    }
    \caption{MNAR}
    \label{fig:dag-mnar}
  \end{subfigure}

  \caption{Missingness mechanisms for $A$, with $X$ and $Y$ fully observed and $R$ indicating whether $A$ is observed. Dashed arrows may be present or absent.}
  \label{fig:missingness-dags}
\end{figure}

The impact of missing exposures on identification and estimation of $\tau$ is strongly influenced by the missingness mechanism, where researchers commonly categorize missingness into three scenarios. Intuitively, the missingness mechanism determines the extent to which units with missing exposures differ from units with exposure measurements available. In the most ideal scenario, the exposure is missing \textit{completely at random} (MCAR), $R\indep(\bX,A,Y)$ (Figure~\ref{fig:dag-mcar}), so that complete-cases are a  representative subset of the study population. In an EHR study, this could occur if a randomly selected subset of exposure records were unavailable because of a technical extraction problem unrelated to patient characteristics, exposure status, or outcome. 

The exposure is said to be \textit{missing at random} (MAR) given  the covariates and outcome if $R \indep A \mid (\bX, Y)$. A more restrictive scenario is when the exposure is  MAR given only the observed covariates if $R\indep A\mid \bX$. MAR given $(\bX,Y)$ allows exposure observation to depend on $Y$, even though $Y$ occurs after exposure. For example, to reduce data collection cost, a two-phase survey may assess past occupational exposure in only a subset of participants, selecting those with the disease with higher probability than those without it. The probability of observing exposure would therefore depend on the outcome. Observation depending only on $\bX$ corresponds to Figure~\ref{fig:dag-mar-x}, while additional dependence on $Y$ is illustrated in Figure~\ref{fig:dag-mar-xy}. Outcome-dependent exposure observation has important implications for methods which address missingness, which we discuss in Section \ref{sec:methods-for-mar}. Throughout, we additionally assume positive exposure observation probabilities so that $\PP(R=1 \mid \bX, Y)>0$.

The final, most challenging scenario occurs when $A \not \indep R \mid (\bX, Y)$, so that $A$ is still predictive of whether it is missing even after conditioning on the always-observed covariates and outcomes. In this setting, the exposure is said to be missing \textit{not at random} (MNAR). This MNAR mechanism is illustrated in Figure~\ref{fig:dag-mnar}. This could occur if patients' willingness to report an exposure depended on their exposure status, even after accounting for recorded characteristics and outcomes.

\subsection{Estimation with Complete Exposure Information}

We briefly outline the estimation of $\tau$ in the ideal scenario where complete data $O^\text{C}$ is available, since the missing data methods we discuss can  all be viewed as imputing different aspects of the complete data estimator. An increasingly standard approach to estimating $\tau$ is through augmented inverse probability weighting (AIPW). Specifically, defining the complete data contributions
\begin{align*}
   H(O^\text{C}_i;m,g) =    m(\bX_i,1) -  m(\bX_i,0) + \left( \frac{A_i}{ g(\bX_i)} - \frac{1-A_i}{1- g(\bX_i)}\right)\{Y_i - m(\bX_i,A_i) \},
\end{align*}
the AIPW estimator is
\[
\hat \tau^\text{AIPW} = \sumn H(O^\text{C}_i \ ;   \widehat{m}, \widehat{g}),
\]
where the estimated regression and propensity score functions $\widehat{m}(\bX,A)$ and $\widehat{g}(\bX)$ can be obtained either through a standard parametric estimation procedure or a predefined machine learning algorithm, and in turn we leave their specific forms generic. AIPW combines predicted outcomes under each exposure level with an inverse propensity-weighted correction based on observed residuals of those predictions. This estimator is consistent if either the outcome regressions or the propensity score is consistently estimated \citep{BangRobins2005}, and can accommodate flexible machine learning methods for estimating these functions while still allowing for valid statistical inference \citep{naimi2023challenges}. Unfortunately, $\hat \tau^\text{AIPW}$ is infeasible in the presence of missing data, since the above contributions $H(O^\text{C} ; \widehat{m}, \widehat{g})$ are a function of $A$, and in turn themselves missing whenever the exposure is missing.

\subsection{Complete-Case Analyses}

Before discussing approaches that directly address missing exposure information, it is illustrative to consider issues that can arise in complete-case (CC) estimation of $\tau$, which remains one of the most common analytical approaches for addressing missing data \citep{baker2025systematic}.  CC AIPW  emulates the above full-data AIPW analysis by restricting estimation to the subset of units with $R=1$, fitting $\widehat{m}_\text{CC}(\bX,a) = \hat \E(Y \mid A=a,\bX,R=1)$ and $\widehat{g}_\text{CC} = \hat \PP(A=1\mid \bX,R=1)$ on this subset and forming the analogous estimate
\[\widehat{\tau}_{\mathrm{CC}}=\frac{1}{n_R}\sum_{i:R_i=1}H(O^\text{C}_i;\widehat m_{\mathrm{CC}},\widehat g_{\mathrm{CC}}),\]
where $n_R=\sum_{i=1}^{n}R_i$ denotes the number of units with an observed exposure. When exposures are MCAR, complete-case have the same distribution as the target population. Therefore, complete-case AIPW can consistently estimate $\tau$,
although discarding incomplete records generally reduces precision. 

When exposures are not MCAR, however, two distinct problems can arise. \textit{First}, selection bias among complete cases can change the corresponding outcome regression and treatment propensity score function for this subpopulation. In particular, $m_\text{CC}(\bX,a)$ and $g_\text{CC}(\bX) $ are not guaranteed to equal their full-data counterparts. Importantly, this problem is separate from model misspecification; nuisance learners that consistently estimate $m_\text{CC}(\bX,a)$ and $g_\text{CC}(\bX)$ will generally be inconsistent for their full-data counterparts when exposures are not MCAR. \textit{Second}, the AIPW contributions are averaged over the complete-case covariate distribution. When $\widehat{m}_\text{CC}$ and $\widehat{g}_\text{CC}$ are consistently estimated, the complete-case estimator then targets the complete-case analogue of the G-formula estimand
\[
\mathbb{E}\{m_\text{CC}(\bX,1)-m_\text{CC}(\bX,0)\mid R=1\},
\]
which may differ from $\tau$ when exposure availability ($R$) varies with $\bX$ and the conditional average treatment effect is heterogeneous \citep{ross2020complete}. In turn, the double robustness of complete-case AIPW does not generally protect against bias induced by missing exposure data. Importantly, these sources of bias arise regardless of whether exposures are MAR or MNAR, since the CC estimator does not use measured variables to account for missingness.

\section{Methods that Account for Missing Data}
\label{sec:methods-for-mar}

A wide array of methods have been developed to address the above sources of bias in CC analyses. These approaches can be understood as imputing missing components of the  full-data AIPW estimator, though they differ in components they impute and the complexity of their implementation. We briefly outline these classes of approaches below, where throughout we assume exposures are MAR given $(\bX, Y)$ and discuss adjustments for when the researcher is willing to assume exposures are MAR given $\bX$.

A key consideration in implementing the approaches below is whether to include the outcome $Y$ when modeling exposure missingness or imputing missing exposures. This choice can be  unclear given the broader emphasis in the propensity score literature on separating the study design and analysis phases \citep{rubin2007design}, and the related emphasis on cutting feedback between outcome and treatment modeling in Bayesian causal inference \citep{zigler2013model}, which would seem to imply that $Y$ should not be included in either model. We consider this choice for each of the methods introduced below.

\paragraph{Single Imputation} \
Single imputation is motivated by the observation that when exposures are MCAR or MAR, $\PP(A=1\mid \bX,Y,R=1)=\PP(A=1\mid \bX,Y)$, implying a model fitted among complete-cases can  be used to predict missing exposures. A particular variant---stochastic single imputation---draws a value $\hat A$ from the fitted distribution $\hat \PP(A=1 \mid \bX, Y, R=1)$ for each individual with $R=0$. After imputation, the outcome regression $m(\bX,A)$ and exposure propensity score model $g(\bX)$ are fitted using the completed dataset. One can then construct a final AIPW estimator which averages the resulting contributions $H(O^\text{C}_i;\widehat m,\widehat g)$.
 
Unlike the exposure propensity score $g(\bX)$, the exposure imputation model generally needs to include $Y$ to preserve the exposure-outcome relationship, regardless of whether $A$ is MAR given $\bX$ alone \citep{MoonsEtAl2006}. Importantly, drawing imputed exposures as a function of only $\bX$ artificially renders the imputed $\hat A$ independent of $Y$ within covariate strata, which can attenuate estimated exposure-outcome associations. 

The validity of single imputation depends on both the imputation and analysis models. Treating singly-imputed exposure values as observed can underestimate standard errors by ignoring uncertainty resulting from the missing data \citep{SterneEtAl2009}.

\paragraph{Multiple Imputation}
Motivated by single imputation's difficulty in properly propagating uncertainty from the imputation model, multiple imputation addresses missing exposure data by generating $M$ completed datasets using a \textit{stochastic} imputation procedure, rather than a single imputed dataset. AIPW is applied separately to each completed dataset, and the resulting estimates and estimated variances are combined using Rubin's rules \citep{Rubin1987}. The pooled variance combines within-imputation variance and between-imputation variability:
\[ T=\overline U+(1+1/M)B\]
where $\overline U$ is the mean estimated within-imputation variance and $B$ is the between-imputation variance. A choice of $M=10$ has been used in causal inference simulations \citep{LeyratEtAl2021}, although larger numbers may be needed to obtain stable standard error estimates \citep{vonHippel2020}. Since multiple imputation is essentially a generalized version of stochastic single imputation where $M>1$, including $Y$ in the exposure imputation model is similarly required for consistency regardless of whether exposures are MAR given only $\bX$.

\paragraph{Multiply-Robust Estimation} Recall that the ideal, but unfeasible, full-data AIPW estimator averages the contributions $H(O^\text{C};\widehat{m},\widehat{g})$ for all $n$ units. Importantly, these contributions are  unobserved for units with $R=0$, since they depend on the partially unobserved exposures. Motivated by broader work on enabling model-robust estimation in the presence of missing data \citep{RobinsRotnitzkyZhao1994}, the multiply-robust (MR) approach imputes these contributions \textit{directly}, rather than the missing exposure.  For given nuisance functions $m$ and $g$, define $q_{m,g}(\bX,Y) = \E\{H(O^\text{C};m,g)\mid \bX,Y\}.$ Rather than impute the missing components of the complete-data AIPW contribution $H(O^\text{C}; m, g)$, the MR approach directly imputes these contributions  as a function of the always-observed $\bX$ and $Y$---essentially treating $H(O_i^\text{C}); \widehat m, \widehat g)$ as a \textit{pseudo-outcome}---using variables observed for every individual. Let $\pi(\bX,Y)=\PP(R=1\mid \bX,Y)$ denote the exposure observation probability. The MR estimator is
\[
\widehat{\tau}_{\mathrm{MR}}
=\frac{1}{n}\sum_{i=1}^{n}
\left[
\widehat{q}(\bX_i,Y_i)
+\frac{R_i}{\widehat{\pi}(\bX_i,Y_i)}
\left\{
H(O^\text{C}_i;\widehat m,\widehat g)
-\widehat{q}(\bX_i,Y_i)
\right\}
\right].
\]
Noting that the population-level mean of $H(O^\text{C};m,g)$ equals $\tau$ under consistency, positivity and unconfoundedness, $\hat \tau_\text{MR}$ can be viewed as an AIPW estimator for the mean of the partially missing $H(O^\text{C}; m,g)$: analogous to complete-data AIPW, $\hat \tau^\text{MR}$ augments regression predictions for the pseudo-outcome with inverse probability weighted residuals from that regression (for units with observed exposure). Importantly, $\hat \tau_\text{MR}$ can be shown to be consistent for $\tau$ so long as (i) one of $\widehat{m}$ and $\widehat{g}$ is consistent, and (ii) one of $\hat q$ and $\hat \pi$ is consistent \citep{levis2024double}.

When exposure observation only depends on $\bX$, $\pi(\bX,Y) = \tilde \pi(\bX) = \PP(R=1\mid \bX)$, implying both the observation propensity score and pseudo-outcome regression can use $\bX$ alone, unlike single and multiple imputation which always require including $Y$. Although $Y$ is not required in such settings, in settings where the exposure is strongly associated with $Y$, including the outcome in both nuisance models can improve efficiency of the MR estimator, akin to including variables predictive of the outcome alone in complete-data AIPW estimation \citep{shen2014inverse}.


\section{Simulation}
\label{sec:simulation}

As discussed in Section \ref{sec:missing-exp}, missing exposures can arise under a range of underlying mechanisms, and it can be difficult to determine how those mechanisms influence the relative performance of the above methods. To this end, in this Section we perform a set of simulation studies to illustrate settings where missing exposures can induce bias in complete-case analyses, and how the above approaches that account for missingness perform across these settings.

We compare MCAR, MAR, and MNAR mechanisms, examine exposure--covariate interactions, and assess the consequences of excluding or including the outcome when accounting for missing exposures. An additional experiment using nonlinear transformations of the covariates is presented in Appendix~\ref{app:nonlinear}.

\subsection{Setup}
Our main simulation adapted the design of \cite{shi2025causal}, who developed and evaluated weighting and triply robust estimators for causal inference with exposures missing at random. Specifically, we generated three independent covariates $X_1,X_2,X_3\sim N(0,1)$, and then sequentially generated exposure and outcome according to
\begin{align}
  \operatorname{logit}\Pr(A=1\mid \bX)
    &= -0.2+0.9X_1+1.0X_2+0.8X_3, \\
  \operatorname{logit}\Pr(Y=1\mid A,\bX)
    &= 0.7+A+0.5X_1+0.9X_2+0.7X_3 \notag\\
    &\quad+\gamma_{Y,AX}A(X_1+X_2+X_3).
  \label{eq:main_dgp}
\end{align}
Above, $\gamma_{Y,AX}$ controls the level of exposure--covariate interaction (i.e., effect heterogeneity) in the outcome model. We generated the exposure observation indicator from
\begin{equation}
\operatorname{logit}\Pr(R=1\mid\bX,Y,A)
=
\gamma_{R,0}
+
\delta_R\left(
0.6X_1+0.7X_2+0.8X_3+\gamma_{R,Y} Y
\right)
+
\gamma_{R,A} A.
\end{equation}
We set $\delta_R=0$ and $\gamma_{R,A}=0$ under MCAR. Under MAR, we set $\delta_R=1$ and $\gamma_{R,A}=0$, whereas under MNAR we set $\delta_R=1$ and $\gamma_{R,A}=2$, $\gamma_{R,Y}=0.5$. The intercept $\gamma_{R,0}$ was calibrated separately for each scenario to target an expected missing exposure proportion $\PP(R=0)$ of 40\%.

Throughout our main exercise, we fix $n=2000$. We first compared the four exposure observation settings outlined in Figure \ref{fig:missingness-dags}---MCAR, MAR given $\bX$, MAR given $(\bX, Y)$, and MNAR---while fixing $\gamma_{Y,AX}=0$.  We additionally examined exposure--covariate interaction by varying $\gamma_{Y,AX}$ across $\{0,0.5,1\}$ under MCAR (Figure~\ref{fig:dag-mcar}) and MAR given $(\bX,Y)$ (Figure~\ref{fig:dag-mar-xy}). In the latter mechanism, $\gamma_{R,Y}$ was fixed at 0.5.

Finally, we compared excluding and including $Y$ when accounting for missing exposures under MAR given $\bX$ ($\gamma_{R,Y}=0$) (Figure~\ref{fig:dag-mar-x}) and MAR given $(\bX,Y)$ ($\gamma_{R,Y}=2$)(Figure~\ref{fig:dag-mar-xy}), with $\gamma_{Y,AX}=0$. For single and multiple imputation, the specification excluding $Y$ used $\bX$ as predictors, whereas the specification including $Y$ used $(\bX,Y)$. Each displayed estimator combination contained 500 valid Monte Carlo replications.

\paragraph{Analysis approaches}
In estimating the outcome regression we specified a logistic model,
\[\operatorname{logit}m(A,\bX)=\beta_0+\beta_A A+\sum_{j=1}^{3}\beta_jX_j.\]
In the interaction specification, it additionally included $\beta_I A(X_1+X_2+X_3)$, with one common interaction coefficient. $\widehat m(\bX,1)$ and $\widehat m_0(\bX)$ were obtained from predictions under $A=1$ and $A=0$. exposure propensity models were main-effects logistic regressions of $A$ on $\bX$ throughout.

Except in the specific setting excluding $Y$, exposure imputation models used $X_1,X_2,X_3$, and $Y$ as predictors. Under the interaction scenario, these models additionally included $YX_1,YX_2,YX_3$, with separate coefficients. For single and multiple imputation, the interaction specification included interaction terms in both the exposure imputation model and the outcome model used in the following AIPW analysis. We estimated the exposure observation model $\pi(\bX, Y)$ with a main-effects logistic regression of $R$ on $(\bX,Y)$. Outcome and exposure propensity models were fitted using observation weights $R/\widehat\pi(\bX,Y)$. Exposure-specific pseudo-outcomes were regressed on $(\bX,Y)$ using generalized linear models among observed exposures.

We used the \texttt{AIPW} \citep{zhong2021aipw} package to implement complete-data, complete-case and imputation-based AIPW, where the \texttt{mice} package \citep{van2011mice} was additionally used in implementing multiple imputation. The \texttt{drcmd} package \citep{BarnatchezDesRoches2026} was used to implement the MR estimator. Complete implementation details are provided in Appendix \ref{app:implementation_settings}. Performance was assessed using relative bias (\%), root mean squared error (RMSE), and 95\% confidence interval coverage. Definitions of these measures and additional details are provided in Appendix~\ref{app:performance-measures}.

\subsection{Missingness mechanisms and exposure--covariate interaction}

\paragraph{Exposure observation mechanisms.}

\begin{figure}[H]
\centering
\includegraphics[width=\linewidth]{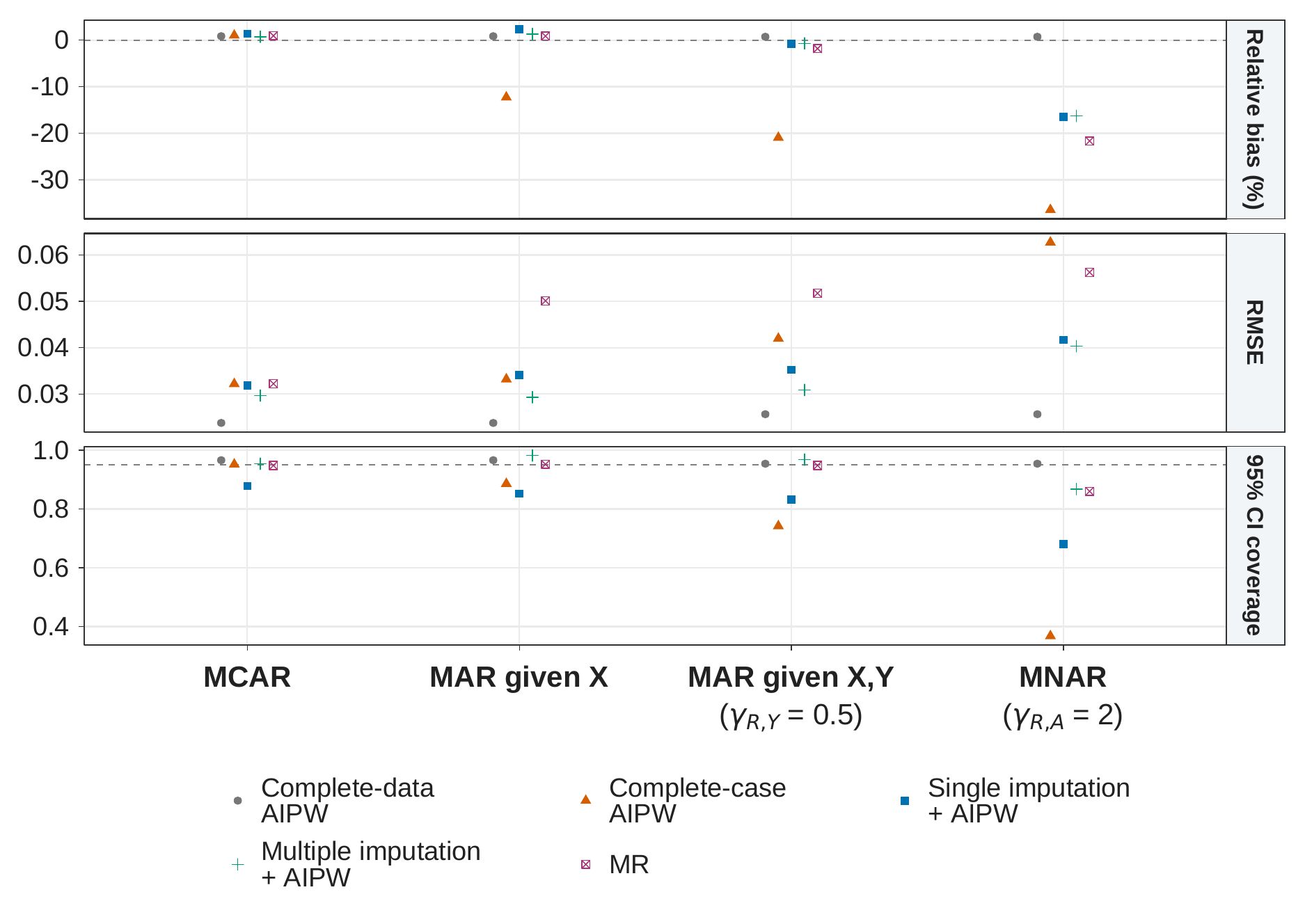}
\caption{\textbf{Performance under MCAR, MAR, and MNAR exposure missingness.} The MAR settings allow exposure observation to depend on covariates alone or on both covariates and the outcome ($\gamma_{R,Y}=0.5$). In the MNAR setting, observation additionally depends on the exposure itself after conditioning on $(\bX,Y)$, with $\gamma_{R,A}=2, \gamma_{R,Y}=0.5$. All settings use $\gamma_{Y,AX}=0$.
}
\label{fig:missing-exposure-2}
\end{figure}

Figure~\ref{fig:missing-exposure-2} compares MCAR, MAR given $\bX$, MAR given $(\bX,Y)$, and MNAR exposure missingness. Under MCAR, complete-case AIPW had little bias and coverage close to the nominal 95\% level. It performs poorly when exposure observation depends on covariates or the outcome. Relative bias was approximately $-13\%$ under MAR given $\bX$ and $-21\%$ under MAR given $(\bX,Y)$, with coverage decreasing to approximately 88\% and 74\%, respectively. Single and multiple imputation followed by AIPW and the MR estimator had small relative bias under MCAR and the two MAR mechanisms. However, single imputation had coverage below 90\% under all settings. Multiple imputation followed by AIPW and MR maintained coverage close to the nominal level under MCAR and the two MAR mechanisms, although MR had higher RMSE than the other approaches using incomplete exposure data in the MAR settings. Thus, similar point estimation bias did not imply similar precision or confidence interval coverage.

Under MNAR, all approaches using incomplete exposure data exhibited negative relative bias.  Complete-case AIPW had the largest relative bias ($-36\%$) and lowest coverage ($37\%$). Relative bias was approximately $-16\%$ for single and multiple imputation and $-22\%$ for MR. The complete-data benchmark remained nearly unbiased across all mechanisms.

\paragraph{Exposure--covariate interaction.}

\begin{figure}[h!]
\centering
\includegraphics[width=\linewidth]{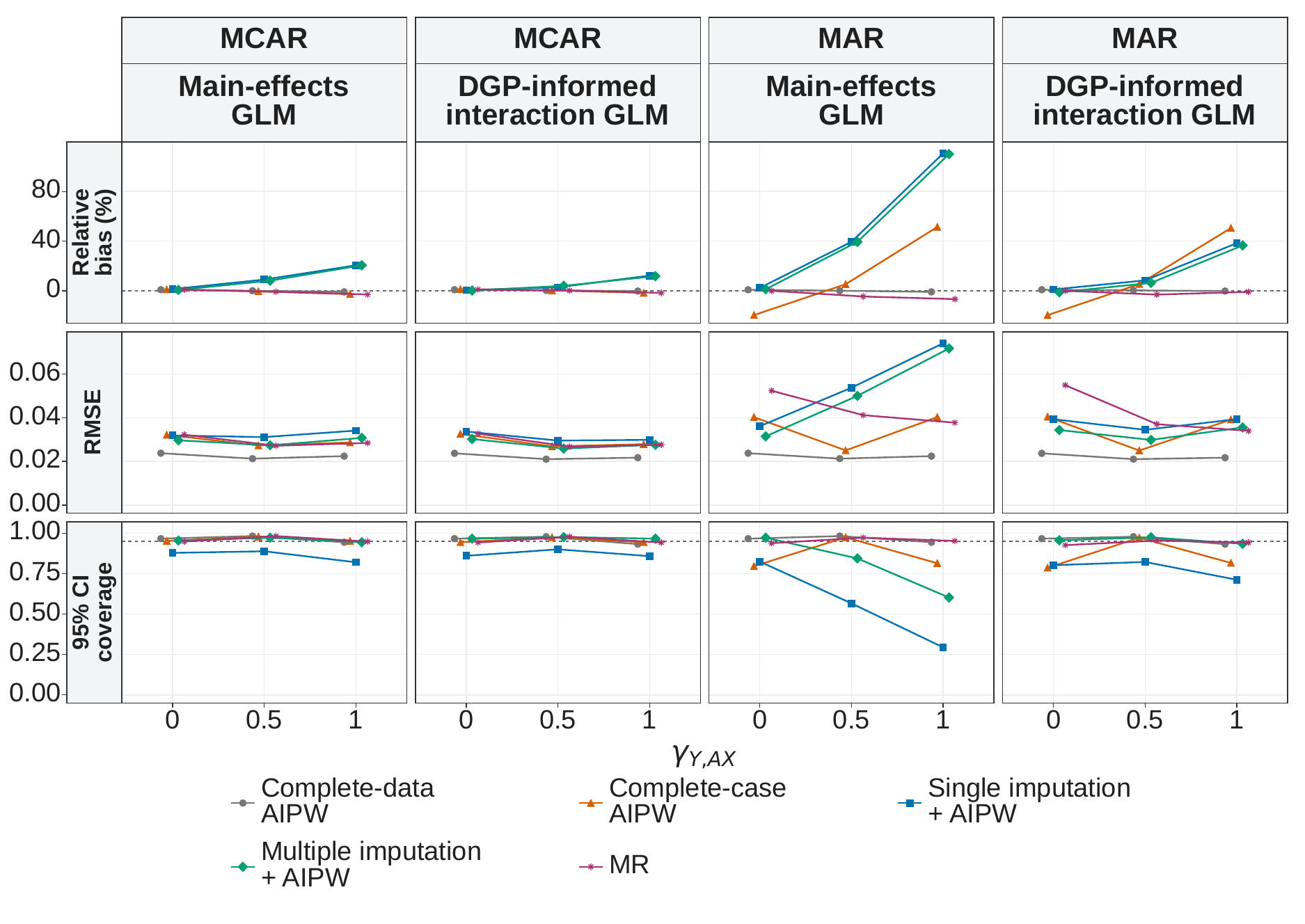}
\caption{\textbf{Performance under MCAR and MAR with varying exposure–covariate interaction.} The coefficient $\gamma_{Y,AX}$ controls the strength of exposure--covariate interaction in the DGP: values of $0$, $0.5$, and $1$ correspond to no, moderate, and strong interaction, respectively. In the MAR scenarios, exposure observation depends on both baseline covariates $\mathbf{X}$ and the outcome $Y$, with $\gamma_{R,Y}=0.5$, and is conditionally independent of exposure $A$ given $(\mathbf{X},Y)$.}
\label{fig:missing-exposure-1}
\end{figure}

Figure~\ref{fig:missing-exposure-1} displays results when varying interaction strength and model specification.
Under MCAR, complete-case AIPW had little bias and coverage close to the nominal level across different interaction strengths. Under the main-effects specification, positive bias in single and multiple imputation followed by AIPW increased with interaction strength. Bias was smaller under the specification that included interactions in both the outcome and exposure imputation models. As expected, single imputation still showed undercoverage under both specifications.

Under MAR, with main effects models bias and RMSE increased for both single and multiple imputation with the degree of exposure-covariate interaction. Coverage also declined to approximately 30\% for single imputation and 60\% for multiple imputation at the strongest interaction setting. Using interaction terms in both the outcome and exposure imputation models reduced bias and RMSE and also improved coverage, although single imputation continued to show undercoverage. 
Complete-case AIPW bias changed from negative to positive as interaction strength increased, while complete-data AIPW and MR maintained small bias and coverage close to 95\% across these settings.

\subsection{The role of outcome information}
\begin{figure}[h!]
\centering
\includegraphics[width=\linewidth]
{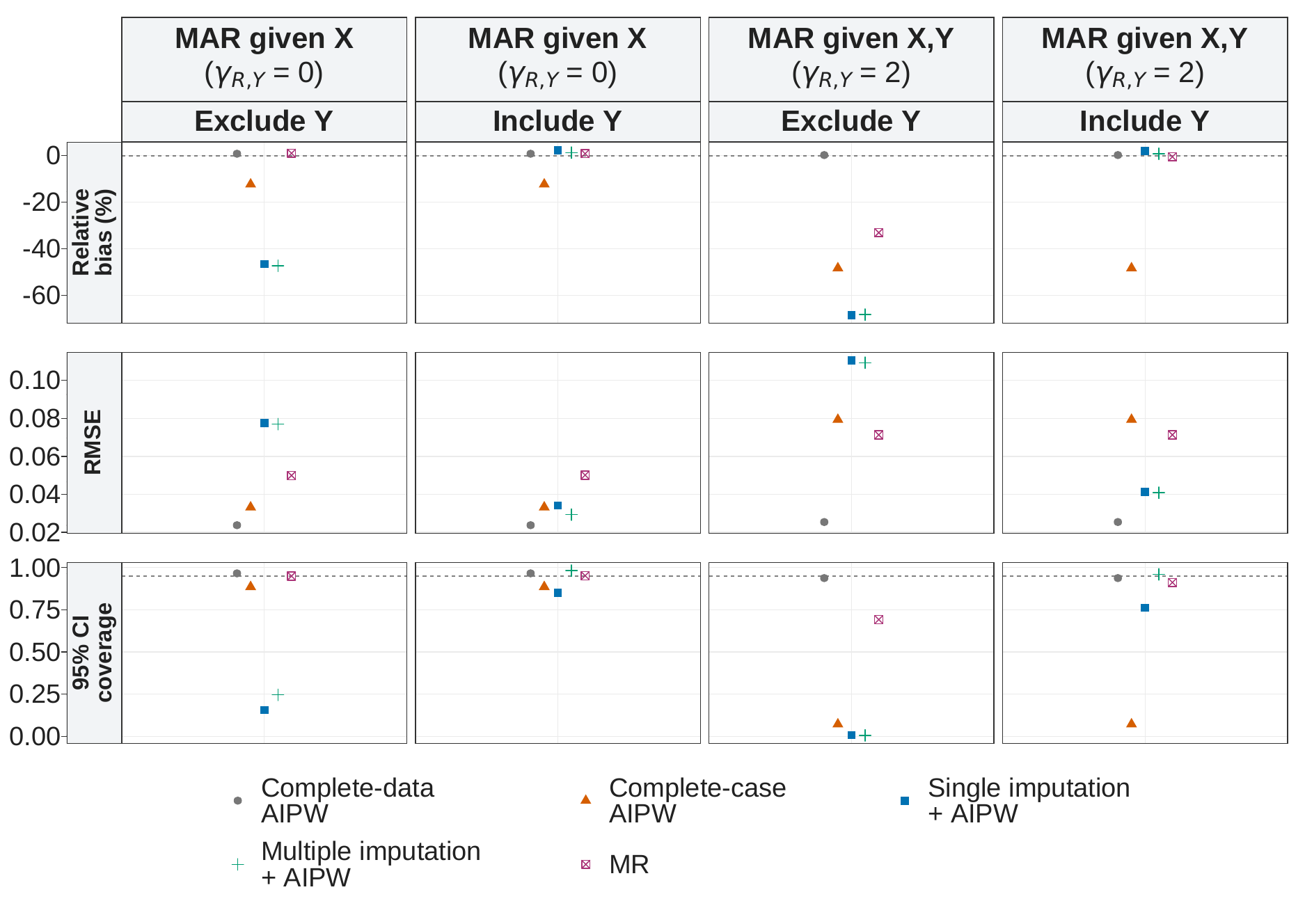}
\caption{
\textbf{Performance when including or excluding $Y$ in models for missing exposures.}
Columns compare exclude-$Y$ and include-$Y$ specifications
under MAR given $\bX$ ($\gamma_{R,Y}=0$) and MAR given $(\bX,Y)$
($\gamma_{R,Y}=2$).
For single and multiple imputation, these specifications
exclude or include $Y$ as a predictor of the missing exposure. For MR, $Y$ is excluded from or included as a predictor in both the exposure observation model and the pseudo-outcome regression.
Complete-data and complete-case AIPW are repeated as reference
analyses within each missingness mechanism.
}
\label{fig:include-exclude-y}
\end{figure}

To examine when outcome information is needed in implementing the above methods, we compared specifications excluding and including $Y$ under MAR given $(\bX,Y)$ and MAR given only $\bX$. Figure \ref{fig:include-exclude-y} displays the results.


Excluding $Y$ from the imputation model produced substantial negative bias for both single and multiple imputation. Relative bias was about $-46\%$ to $-47\%$ under MAR given $\bX$, with coverage below 25\% for both approaches. Under MAR given $(\bX, Y)$, relative bias and coverage were $-68\%$ and below $1\%$ for both imputation approaches, respectively. Importantly, including $Y$ reduced bias to nearly zero and markedly reduced RMSE, with coverage rising to $96-98\%$ for multiple imputation and $76-85\%$ for single imputation. Notably, under MAR given $\bX$ MR had little bias and near-nominal coverage regardless of specification, whereas under MAR given $(\bX, Y)$ including $Y$ reduced relative bias from $-33\%$ to $-0.5\%$ and increased coverage from $69\%$ to $91\%$.

These results illustrate that the variables needed for imputation are not determined only by the variables governing missingness. Even when exposure observation depends only on $\bX$, the outcome can contain information about the missing exposure conditional on $\bX$. Omitting this information can distort the relationship between exposure and outcome in the completed data \citep{MoonsEtAl2006}. Including $Y$ in imputation can help address a different issue from including covariates in the propensity score model for confounding adjustment.

\subsection{Takeaways}
Our simulations suggest three practical considerations when choosing an approach to missing exposure data.

\paragraph{1. Double robustness does not automatically protect against missing exposure bias.}

Complete-case AIPW performed well under MCAR but showed bias and reduced coverage in several MAR settings. In practice, the missingness mechanism is usually unknown, so the decision to use complete-case analysis should be informed by subject-matter knowledge, the study design, and how exposure information was collected.

Under MCAR, complete-case form a random subset of the study population, supporting estimation of the population ATE. Validity can also extend beyond MCAR, as if exposure observation depends only on baseline covariates, complete-case AIPW can remain valid for the population ATE when the conditional risk difference is constant across those covariates \citep{ross2020complete}. However, if the conditional risk difference varies across covariate groups, changes in the representation of these groups among complete-case can make their average effect differ from the population ATE. These conditions should be justified rather than assumed. For example, in a cohort study with unanswered exposure questions, knowledge of questionnaire administration and reasons for nonresponse can help assess whether restricting the analysis to respondents is defensible. Even when complete-case analysis is valid, it
can lose precision by discarding observed
covariates and outcomes from participants with
missing exposures. When MAR is plausible,
methods such as multiple imputation \citep{WhiteEtAl2011} and MR estimation can use this information. Their performance depends on the models used to recover missing exposures or AIPW contributions.

\paragraph{2. Including the observed outcome is a reasonable default when accounting for missing exposures.}
For stochastic single and multiple imputation, both theory and our simulations suggest that the outcome should be included in the exposure imputation model, regardless of whether the outcome influences exposure availability \citep{feldman2026outcome}. For MR estimation, adjustment for the outcome is only necessary when exposure missingness depends on $Y$, however our simulations suggest that choosing whether to adjust for $Y$ when exposures are MAR given only $\bX$ made little difference to bias or coverage. Even in such settings, adjusting for the outcome can provide efficiency gains provided the outcome provides information on the exposure after covariate adjustment \citep{Kennedy2020}, which will typically be the case when the exposure has a causal effect on the outcome. 

\paragraph{3. Account for missing exposures in both estimation and inference.}

Although single imputation occasionally yielded little bias and relatively low RMSE, its inability to propagate imputation uncertainty generally yielded confidence intervals with below-nominal coverage. MI and MR estimation provided theoretically principled alternatives to  complete-case analysis, though our simulations suggest that their performance is strongly tied to model specification and the underlying data-generating process. For either approach, examining how effect estimates and confidence intervals change under reasonable model specifications can help assess the sensitivity of the conclusions. The missingness assumption also matters, as neither MI nor MR eliminated bias in our MNAR experiment.

\section{Discussion}

Missing exposure data are common epidemiologic studies, and restricting doubly-robust methods like AIPW to complete cases can bias resulting effect estimates. In this paper, we provided clarity on approaches that adapt complete-data AIPW to account for missing exposures. Our simulations showed that complete-case bias varied with the missingness mechanism and treatment effect heterogeneity, and that methods yielding similarly small bias could differ substantially in precision and confidence interval coverage.

While we considered multiple approaches for addressing missing exposures, there are additional approaches which address missing exposures that can be viewed as extensions of AIPW \citep{shi2025causal,barnatchez2025flexible,sui2026efficient}. Additionally, for simplicity we  considered cross-sectional settings where the observed data is sampled directly from the target population of interest. Applying the methods considered here and these broader approaches to more complex study designs may require additional adaptations, such as sampling weights and design based variance estimation for complex surveys \citep{Lumley2004}, methods accounting for within cluster dependence in clustered or repeatedly measured data \citep{LiangZeger1986}, and assessment of residual temporal autocorrelation in environmental time-series analyses \citep{BhaskaranEtAl2013}.

\paragraph{Code Availability} Code to reproduce results from our simulation study is available at \url{https://github.com/ThisisMomoW/missing-exposure-simulation}.

\paragraph{Funding} This work was partially supported by an iLead project from Eli Lilly \& Co, as well as  National Institutes of Health (NIH) grant T32HL139426 and the Patient-Centered Outcomes Research Institute (PCORI) grant ME-2025C2-45978, .

\bibliographystyle{apalike}
\bibliography{sources}

@article{ZhangEtAl2016,
  author  = "Zhang, Z. and Liu, W. and Zhang, B. and Tang, L. and Zhang, J.",
  title   = "Causal inference with missing exposure information: Methods and applications to an obstetric study",
  journal = "Statistical Methods in Medical Research",
  volume  = "25",
  number  = "5",
  pages   = "2053--2066",
  year    = "2016",
  doi     = "10.1177/0962280213513758"
}

@article{RobinsRotnitzkyZhao1994,
  author  = "Robins, J. M. and Rotnitzky, A. and Zhao, L. P.",
  title   = "Estimation of regression coefficients when some regressors are not always observed",
  journal = "Journal of the American Statistical Association",
  volume  = "89",
  number  = "427",
  pages   = "846--866",
  year    = "1994",
  doi     = "10.1080/01621459.1994.10476818"
}

@article{BangRobins2005,
  author  = "Bang, H. and Robins, J. M.",
  title   = "Doubly robust estimation in missing data and causal inference models",
  journal = "Biometrics",
  volume  = "61",
  number  = "4",
  pages   = "962--973",
  year    = "2005",
  doi     = "10.1111/j.1541-0420.2005.00377.x"
}

@book{Rubin1987,
  author    = "Rubin, D. B.",
  title     = "Multiple Imputation for Nonresponse in Surveys",
  publisher = "John Wiley \& Sons",
  year      = "1987",
  doi       = "10.1002/9780470316696"
}

@article{DondersEtAl2006,
  author  = "Donders, A. R. T. and van der Heijden, G. J. M. G. and Stijnen, T. and Moons, K. G. M.",
  title   = "Review: A gentle introduction to imputation of missing values",
  journal = "Journal of Clinical Epidemiology",
  volume  = "59",
  number  = "10",
  pages   = "1087--1091",
  year    = "2006",
  doi     = "10.1016/j.jclinepi.2006.01.014"
}

@article{shi2025causal,
  title={Causal Inference on Missing Exposure via Robust Estimation},
  author={Shi, Yuliang and Zhu, Yeying and Dubin, Joel A},
  journal={Statistics in Biosciences},
  pages={1--31},
  year={2025},
  publisher={Springer}
}

@misc{BarnatchezDesRoches2026,
  author       = "Barnatchez, K. and DesRoches, G.",
  title        = "{drcmd}: Doubly-Robust Causal Inference with Missing Data",
  year         = "2026",
  note         = "R package version 0.1.0",
  howpublished = "\url{https://github.com/keithbarnatchez/drcmd}"
}

@article{KangSchafer2007,
  author  = "Kang, J. D. Y. and Schafer, J. L.",
  title   = "Demystifying double robustness: A comparison of alternative strategies for estimating a population mean from incomplete data",
  journal = "Statistical Science",
  volume  = "22",
  number  = "4",
  pages   = "523--539",
  year    = "2007",
  doi     = "10.1214/07-STS227"
}

@article{SamoliEtAl2014,
  author = {Samoli, Evangelia and Peng, Roger D. and Ramsay, Tim and Touloumi, Giota and Dominici, Francesca and Atkinson, Richard W. and Zanobetti, Antonella and Le Tertre, Alain and Anderson, H. Ross and Schwartz, Joel and Cohen, Aaron and Krewski, Daniel and Samet, Jonathan M. and Katsouyanni, Klea},
  title = {What is the impact of systematically missing exposure data on air pollution health effect estimates?},
  journal = {Air Quality, Atmosphere and Health},
  year = {2014},
  volume = {7},
  number = {4},
  pages = {415--420},
  doi = {10.1007/s11869-014-0250-2}
}

@article{ShortreedForbes2010,
  author  = "Shortreed, S. M. and Forbes, A. B.",
  title   = "Missing data in the exposure of interest and marginal structural models: A simulation study based on the Framingham Heart Study",
  journal = "Statistics in Medicine",
  volume  = "29",
  number  = "4",
  pages   = "431--443",
  year    = "2010",
  doi     = "10.1002/sim.3801"
}

@article{LeyratEtAl2021,
  author  = {Leyrat, Cl{\'e}mence and
             Carpenter, James R. and
             Bailly, S{\'e}bastien and
             Williamson, Elizabeth J.},
  title   = {Common Methods for Handling Missing Data in
             Marginal Structural Models: What Works and Why},
  journal = {American Journal of Epidemiology},
  year    = {2021},
  volume  = {190},
  number  = {4},
  pages   = {663--672},
  doi     = {10.1093/aje/kwaa225}
}

@article{vonHippel2020,
  author  = {von Hippel, Paul T.},
  title   = {How Many Imputations Do You Need?
             A Two-stage Calculation Using a Quadratic Rule},
  journal = {Sociological Methods \& Research},
  year    = {2020},
  volume  = {49},
  number  = {3},
  pages   = {699--718},
  doi     = {10.1177/0049124117747303}
}

@article{MoonsEtAl2006,
  author  = {Moons, Karel G. M. and
             Donders, Rogier A. R. T. and
             Stijnen, Theo and
             Harrell, Frank E., Jr.},
  title   = {Using the outcome for imputation of missing
             predictor values was preferred},
  journal = {Journal of Clinical Epidemiology},
  year    = {2006},
  volume  = {59},
  number  = {10},
  pages   = {1092--1101},
  doi     = {10.1016/j.jclinepi.2006.01.009}
}

@article{WhiteEtAl2011,
  author  = {White, Ian R. and Royston, Patrick and Wood, Angela M.},
  title   = {Multiple imputation using chained equations:
             Issues and guidance for practice},
  journal = {Statistics in Medicine},
  year    = {2011},
  volume  = {30},
  number  = {4},
  pages   = {377--399},
  doi     = {10.1002/sim.4067}
}

@article{Kennedy2020,
  author  = {Kennedy, Edward H.},
  title   = {Efficient Nonparametric Causal Inference with Missing
             Exposure Information},
  journal = {The International Journal of Biostatistics},
  year    = {2020},
  volume  = {16},
  number  = {1},
  doi     = {10.1515/ijb-2019-0087}
}

@article{TanEtAl2023,
  author  = "Tan, A. L. M. and Getzen, E. J. and Hutch, M. R. and others",
  title   = "Informative missingness: What can we learn from patterns in missing laboratory data in the electronic health record?",
  journal = "Journal of Biomedical Informatics",
  volume  = "139",
  pages   = "104306",
  year    = "2023",
  doi     = "10.1016/j.jbi.2023.104306"
}

@article{SterneEtAl2009,
  author = {Sterne, Jonathan A. C. and White, Ian R. and
            Carlin, John B. and Spratt, Michael and Royston, Patrick and
            Kenward, Michael G. and Wood, Angela M. and
            Carpenter, James R.},
  title = {Multiple imputation for missing data in epidemiological
           and clinical research: Potential and pitfalls},
  journal = {BMJ},
  year = {2009},
  volume = {338},
  pages = {b2393},
  doi = {10.1136/bmj.b2393}
}

@article{Lumley2004,
  author  = {Lumley, Thomas},
  title   = {Analysis of Complex Survey Samples},
  journal = {Journal of Statistical Software},
  year    = {2004},
  volume  = {9},
  number  = {8},
  pages   = {1--19},
  doi     = {10.18637/jss.v009.i08}
}

@article{LiangZeger1986,
  author  = {Liang, Kung-Yee and Zeger, Scott L.},
  title   = {Longitudinal data analysis using generalized linear models},
  journal = {Biometrika},
  year    = {1986},
  volume  = {73},
  number  = {1},
  pages   = {13--22},
  doi     = {10.1093/biomet/73.1.13}
}

@article{BhaskaranEtAl2013,
  author  = {Bhaskaran, Krishnan and Gasparrini, Antonio and
             Hajat, Shakoor and Smeeth, Liam and Armstrong, Ben},
  title   = {Time series regression studies in environmental epidemiology},
  journal = {International Journal of Epidemiology},
  year    = {2013},
  volume  = {42},
  number  = {4},
  pages   = {1187--1195},
  doi     = {10.1093/ije/dyt092}
}

@book{Hernan2025CausalInference,
  author    = {Hern{\'a}n, Miguel A. and Robins, James M.},
  title     = {Causal Inference: What If},
  publisher = {CRC Press / Taylor \& Francis},
  year      = {2020},
  address   = {Boca Raton, FL},
  isbn      = {9781420076165}
}

@article{naimi2023challenges,
  title={Challenges in obtaining valid causal effect estimates with machine learning algorithms},
  author={Naimi, Ashley I and Mishler, Alan E and Kennedy, Edward H},
  journal={American Journal of Epidemiology},
  volume={192},
  number={9},
  pages={1536--1544},
  year={2023},
  publisher={Oxford University Press}
}

@article{zivich2021machine,
  title={Machine learning for causal inference: on the use of cross-fit estimators},
  author={Zivich, Paul N and Breskin, Alexander},
  journal={Epidemiology (Cambridge, Mass.)},
  volume={32},
  number={3},
  pages={393},
  year={2021}
}

@article{ross2020complete,
  title={When is a complete-case approach to missing data valid? The importance of effect-measure modification},
  author={Ross, Rachael K and Breskin, Alexander and Westreich, Daniel},
  journal={American journal of epidemiology},
  volume={189},
  number={12},
  pages={1583--1589},
  year={2020},
  publisher={Oxford University Press}
}

@article{zhong2021aipw,
  title={AIPW: an R package for augmented inverse probability--weighted estimation of average causal effects},
  author={Zhong, Yongqi and Kennedy, Edward H and Bodnar, Lisa M and Naimi, Ashley I},
  journal={American Journal of Epidemiology},
  volume={190},
  number={12},
  pages={2690--2699},
  year={2021},
  publisher={Oxford University Press}
}

@article{van2011mice,
  title={mice: Multivariate imputation by chained equations in R},
  author={Van Buuren, Stef and Groothuis-Oudshoorn, Karin},
  journal={Journal of statistical software},
  volume={45},
  pages={1--67},
  year={2011}
}

@article{levis2024double,
  title={Double Sampling for Informatively Missing Data in Electronic Health Record-Based Comparative Effectiveness Research},
  author={Levis, Alexander W and Mukherjee, Rajarshi and Wang, Rui and Fischer, Heidi and Haneuse, Sebastien},
  journal={Statistics in Medicine},
  volume={43},
  number={30},
  pages={6086--6098},
  year={2024},
  publisher={Wiley Online Library}
}

@article{rubin2007design,
  title={The design versus the analysis of observational studies for causal effects: parallels with the design of randomized trials},
  author={Rubin, Donald B},
  journal={Statistics in medicine},
  volume={26},
  number={1},
  pages={20--36},
  year={2007},
  publisher={Wiley Online Library}
}

@article{baker2025systematic,
  title={A systematic review of reporting and handling of missing data in observational studies using the UNOS database},
  author={Baker, William L and Moore, Timothy E and Baron, Eric and Kittleson, Michelle and Parker, William F and Jaiswal, Abhishek},
  journal={The Journal of Heart and Lung Transplantation},
  volume={44},
  number={3},
  pages={462--468},
  year={2025},
  publisher={Elsevier}
}

@article{barnatchez2025flexible,
  title={Flexible and efficient estimation of causal effects with error-prone exposures: a control variates approach for measurement error},
  author={Barnatchez, Keith and Nethery, Rachel and Shepherd, Bryan E and Parmigiani, Giovanni and Josey, Kevin P},
  journal={Biometrics},
  volume={81},
  number={4},
  pages={ujaf151},
  year={2025},
  publisher={Oxford University Press}
}

@article{sui2026efficient,
  title={Efficient estimation of average causal effect with mismeasured exposures},
  author={Sui, Zijian and Zeng, Min and Zhang, Hong},
  journal={Statistical Methods in Medical Research},
  pages={09622802261506990},
  year={2026},
  publisher={SAGE Publications, Sage UK: London, England}
}

@article{zigler2013model,
  title={Model feedback in Bayesian propensity score estimation},
  author={Zigler, Corwin M and Watts, Krista and Yeh, Robert W and Wang, Yun and Coull, Brent A and Dominici, Francesca},
  journal={Biometrics},
  volume={69},
  number={1},
  pages={263--273},
  year={2013},
  publisher={Oxford University Press}
}

@article{shen2014inverse,
  title={Inverse probability weighting for covariate adjustment in randomized studies},
  author={Shen, Changyu and Li, Xiaochun and Li, Lingling},
  journal={Statistics in medicine},
  volume={33},
  number={4},
  pages={555--568},
  year={2014},
  publisher={Wiley Online Library}
}

@article{feldman2026outcome,
  title={Outcome-Assisted Multiple Imputation of Missing Treatments},
  author={Feldman, Joseph and Reiter, Jerome P},
  journal={Observational Studies},
  volume={12},
  number={1},
  pages={67},
  year={2026}
}

@article{williamson2012doubly,
  author  = {Williamson, E. J. and Forbes, A. and Wolfe, R.},
  title   = {Doubly robust estimators of causal exposure effects with missing data in the outcome, exposure or a confounder},
  journal = {Statistics in Medicine},
  year    = {2012},
  volume  = {31},
  number  = {30},
  pages   = {4382--4400},
  doi     = {10.1002/sim.5643}
}

\clearpage

\appendix

\section{Additional Simulation Details}
\label{app:simulation_details}

\subsection{Calibration and calculation of the population ATE}
\label{app:calibration_ate}
In the main simulations, the exposure observation intercept was calibrated separately for each scenario using 200,000 generated observations to target an expected missing exposure proportion of 40\%. The realized proportion could vary across simulation replications.

Population ATEs were approximated by averaging the conditional risk difference over independent samples of 1,000,000 covariate draws for the main experiments and 2,000,000 draws for the nonlinear experiment. A separate population ATE was used for each interaction setting.

\subsection{Software and implementation settings}
\label{app:implementation_settings}
We used 10-fold outer cross-fitting for AIPW and MR estimation \citep{zivich2021machine}. For imputation-based AIPW, imputation was performed before cross-fitting. Outcome and exposure propensity models were then cross-fitted within each completed dataset.

Single imputation used Bernoulli draws from fitted exposure probabilities. Multiple imputation used 10 completed datasets and 10 iterations.

Estimated exposure propensity scores were bounded to $[0.025,0.975]$. For MR estimation implemented using \texttt{drcmd}, estimated exposure observation probabilities were truncated to the same interval.

\subsection{Performance measures}
\label{app:performance-measures}

For each estimator and scenario, let $S$ denote the number of valid simulation replications, $\widehat{\tau}_s$ the estimated ATE in replication $s$, and $\tau$ the true population ATE for that scenario. Let $[L_s,U_s]$ denote the corresponding 95\% confidence interval. We calculated
\begin{align*}
\operatorname{Relative\ bias}\;(\%)
&=
100 \times \frac{1}{S}
\sum_{s=1}^{S}
\left(\widehat{\tau}_s-\tau\right)/\tau,\\
\operatorname{RMSE}
&=
\left\{
\frac{1}{S}
\sum_{s=1}^{S}
\left(\widehat{\tau}_s-\tau\right)^2
\right\}^{1/2},\\
\operatorname{Coverage}
&=
\frac{1}{S}
\sum_{s=1}^{S}
\mathbf{1}\{L_s\leq\tau\leq U_s\}.
\end{align*}
Relative bias was expressed as a percentage, RMSE was reported on the original ATE scale, and coverage was compared with the nominal level of 0.95.

\section{Additional exercise: Nonlinear experiment}
\label{app:nonlinear}

An additional experiment examined nonlinear relationships in the covariates available to the estimators. Following the transformed covariate construction of ~\cite{KangSchafer2007}, we generated independent $Z_1,Z_2,Z_3\sim N(0,1)$ and recorded
\begin{equation}
  X_1=\exp(Z_1/2),\qquad
  X_2=\frac{Z_2}{1+\exp(Z_1)}+10,\qquad
  X_3=\left(\frac{Z_1Z_3}{25}+0.6\right)^3.
  \label{eq:transformed_covariates}
\end{equation}
The estimators used $X=(X_1,X_2,X_3)$, while the data were generated using
$Z=(Z_1,Z_2,Z_3)$:
\begin{align}
  \operatorname{logit}\Pr(A=1\mid Z)
    &=0.3-0.8Z_1+0.8Z_2-0.25Z_3,\\
  \operatorname{logit}\Pr(Y=1\mid A,Z)
    &=1+2A+0.5Z_1+Z_2+0.5Z_3+\gamma_{Y,Z_1Z_2} Z_1Z_2,\\
  \operatorname{logit}\Pr(R=1\mid Z,Y)
    &=\gamma_{R,0}^{\mathrm{NL}}+0.5Z_1-0.5Z_2+0.25Z_3+0.5Y.
  \label{eq:nonlinear_dgp}
\end{align}
Here $\gamma_{Y,Z_1Z_2}$ controls a covariate--covariate interaction, whereas $\gamma_{Y,AX}$ in the main simulation controls an exposure--covariate interaction. Because the transformation contains the information in $Z$, exposure missingness is MAR conditional on $(\bX,Y)$. The intercept $gamma_{R,0}^{\mathrm{NL}}$ was calibrated to 40\% missing exposure. We crossed $n\in\{500,1000,2000\}$ with $\gamma_{Y,Z_1Z_2}\in\{0,0.5,1\}$ and planned 500 replications per design cell. The comparison presented here fixed $\gamma_{Y,Z_1Z_2}=1$. The complete-data benchmark used the same recorded covariates and learner specification as the corresponding incomplete-data analyses.

In this nonlinear experiment, outcome regressions used main-effects logistic models in the recorded covariates under the generalized linear model (GLM) specification and random forests under the random forest (RF) specification. For the MR approach, pseudo-outcome regressions used Gaussian identity-link GLMs and regression forests, respectively. Exposure propensity and exposure observation models remained main-effects logistic regressions. For imputation-based approaches, exposure imputation used logistic models or random forests, with $(X,Y)$ as predictors. RF single imputation used Bernoulli draws from predicted probabilities, whereas RF multiple imputation used terminal-node donor sampling through \texttt{mice}.

\paragraph{Results.}

\begin{figure}[H]
\centering
\includegraphics[width=\linewidth]{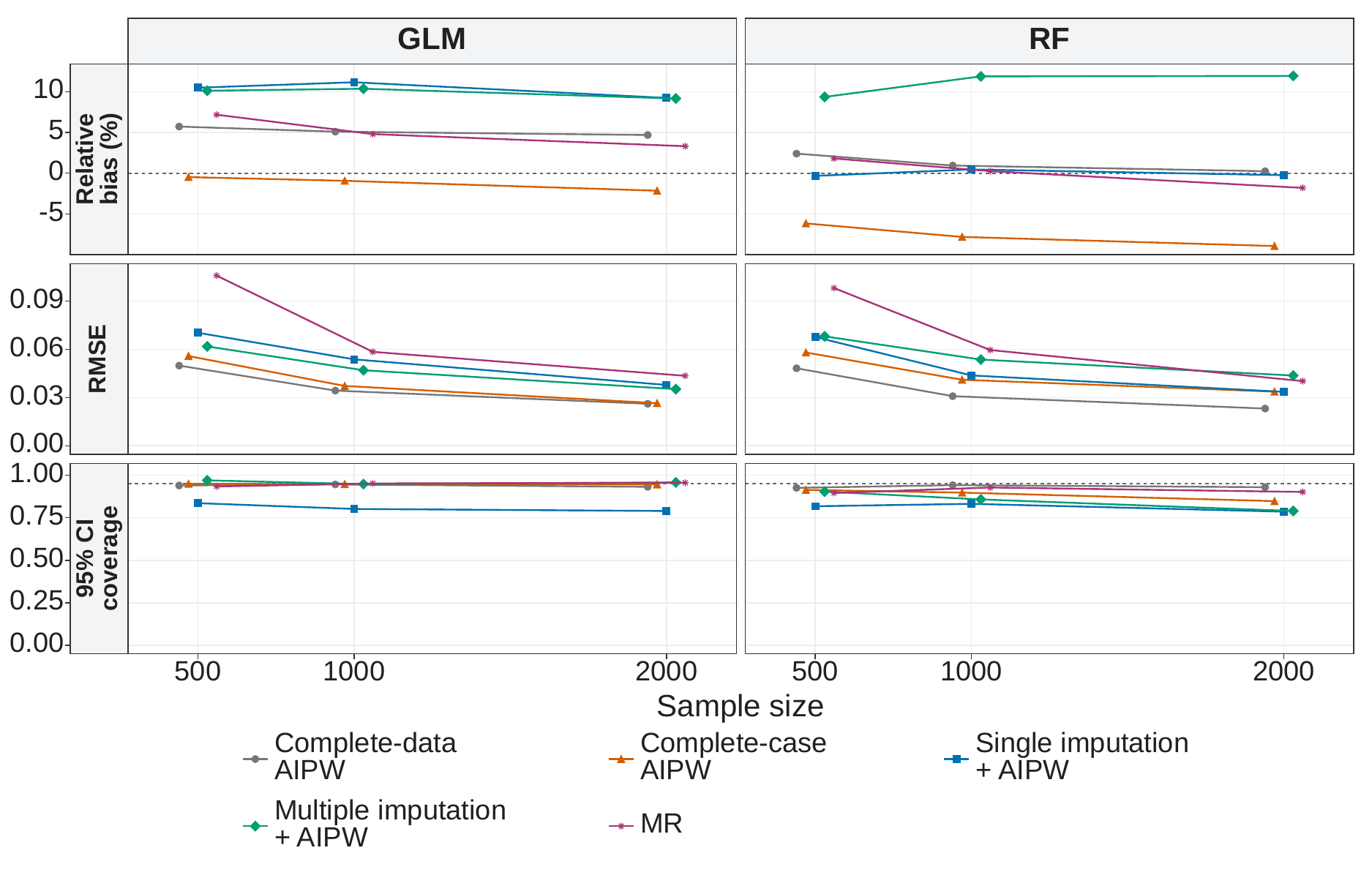}
\caption{\textbf{Performance under a nonlinear DGP using GLM and random forest nuisance estimation.} }
\label{fig:missing-exposure-3}
\end{figure}

Figure~\ref{fig:missing-exposure-3} compares GLM and RF specifications in the nonlinear transformed covariate experiment. Under both specifications, RMSE generally decreased with larger sample size, but increasing sample size did not consistently reduce bias or improve coverage. Random forest reduced the bias of complete-data AIPW and single imputation followed by AIPW compared to their performance under GLM. Nevertheless, single imputation continued to show marked undercoverage under RF. For MR, RF also produced small bias, but its RMSE was similar to that under GLM and its coverage remained below 95\%.

RF did not benefit every approach. Multiple imputation kept positive bias under both specifications, and its RF version had lower coverage than its GLM version. Complete-case AIPW had larger negative bias under RF. For the imputation approaches, both the exposure imputation model and the following AIPW outcome regression changed between columns. Therefore, their differences reflect the combined change in these models. Overall, flexible nuisance estimation improved some point estimates but did not uniformly improve RMSE or confidence interval coverage. For MR, both the outcome and pseudo-outcome regressions changed, while remaining logistic exposure propensity and exposure observation models.

\end{document}